# Radiogenomic biomarkers for immunotherapy in glioblastoma: A systematic review of magnetic resonance imaging studies

Prajwal Ghimire[●], Ben Kinnersley[●], Golestan Karami[●], Prabhu Arumugam, Richard Houlston[●], Keyoumars Ashkan[●], Marc Modat[●], and Thomas C. Booth

All author affiliations are listed at the end of the article

Corresponding Author: Thomas C. Booth, FRCR PhD, School of Biomedical Engineering and Imaging Sciences, King's College London, London, UK (thomas.booth@kcl.ac.uk).

## Abstract

**Background.** Immunotherapy is an effective "precision medicine" treatment for several cancers. Imaging signatures of the underlying genome (radiogenomics) in glioblastoma patients may serve as preoperative biomarkers of the tumor-host immune apparatus. Validated biomarkers would have the potential to stratify patients during immunotherapy clinical trials, and if trials are beneficial, facilitate personalized neo-adjuvant treatment. The increased use of whole genome sequencing data, and the advances in bioinformatics and machine learning make such developments plausible. We performed a systematic review to determine the extent of development and validation of immune-related radiogenomic biomarkers for glioblastoma.

**Methods.** A systematic review was performed following PRISMA guidelines using the PubMed, Medline, and Embase databases. Qualitative analysis was performed by incorporating the QUADAS 2 tool and CLAIM checklist. PROSPERO registered: CRD42022340968. Extracted data were insufficiently homogenous to perform a meta-analysis.

**Results.** Nine studies, all retrospective, were included. Biomarkers extracted from magnetic resonance imaging volumes of interest included apparent diffusion coefficient values, relative cerebral blood volume values, and image-derived features. These biomarkers correlated with genomic markers from tumor cells or immune cells or with patient survival. The majority of studies had a high risk of bias and applicability concerns regarding the index test performed.

**Conclusions.** Radiogenomic immune biomarkers have the potential to provide early treatment options to patients with glioblastoma. Targeted immunotherapy, stratified by these biomarkers, has the potential to allow individualized neo-adjuvant precision treatment options in clinical trials. However, there are no prospective studies validating these biomarkers, and interpretation is limited due to study bias with little evidence of generalizability.

## Key Points

- There are few studies that aim to develop or validate immune-related radiogenomic biomarkers for glioblastoma.
- Radiological biomarkers of key components of the tumor-host immune apparatus have been developed based on apparent diffusion coefficient values, cerebral blood volume values, or radiomics.

Radiogenomics focuses on the relationship between genomics and imaging phenotypes and is increasingly being applied in the research setting to characterize tumors which can be heterogeneous. Characterization might be useful to determine an individual's likelihood of disease progression or immune responsiveness.[1–5] Due to their infiltrative nature, diffuse gliomas typically have a very poor prognosis with the most common type glioblastoma, having a median

## Importance of the Study

We present the first systematic review of immune-related radiogenomic biomarker studies for glioblastoma. Radiological biomarkers of the tumor-host immune apparatus based on apparent diffusion coefficient values, cerebral blood volume values, and image-derived features including VASARI (Visually AcceSAble Rembrandt Images) and more complex radiomics have been developed within the last decade. The summarized evidence provides a basis to further develop and validate future immune-related radiogenomic biomarkers. If validated, these biomarkers have the potential to be further utilized for patient stratification during immunotherapy clinical trials for glioblastoma.

overall survival of only 14.6 months despite standard-of-care treatment (which generally comprises surgery with maximal safe tumor resection, followed by radiotherapy with concomitant and adjuvant temozolomide chemotherapy).[6,7] Recent immunotherapy trials have shown that a subgroup of glioblastoma patients benefit from immune checkpoint inhibitors.[8–10] Furthermore, in a randomized multicenter trial of recurrent glioblastoma, anti-programmed cell death protein-1 (PD-1) neoadjuvant immunotherapy has shown survival benefit.[11] The challenge, however, is that the majority of patients in these studies have shown poor response to immunotherapy, attributable to the immunosuppressive tumor microenvironment (TME) with limited presence of immune cell populations. Current immunotherapies such as PD-1/PD-L1 inhibitors and chimeric antigen receptor T-cell therapy depend on the presence of these tumor-infiltrating lymphocytes within the TME, but these constitute only 10%–15% of all tumor-associated leukocytes.[12,13] In addition, PD-1 expression in human glioma tissues is relatively low as compared to other cancers and is heterogeneous.[14] Despite these challenges, there has been an increased interest in tumor-host immune apparatus target identification in glioblastoma.[9,11] One such area of interest has been to identify preoperative imaging biomarkers that can stratify patients for neo-adjuvant treatment after diagnostic magnetic resonance imaging (MRI). Early and noninvasive diagnosis and treatment therefore has the potential to improve patient quality of life and prolong survival. Noninvasive biomarkers monitoring immunotherapy may also improve patient care.[1–5]

Herein we systematically reviewed 9 studies that developed and validated MRI biomarkers that have the potential to be used, or have been used, for glioblastoma immunotherapy. The primary objective was to analyze immune-related radiogenomic biomarkers. The secondary objective was to highlight alternative methods to develop immunotherapy biomarkers which were not radiogenomic.

## Materials and Methods

We performed a systematic review (registered in PROSPERO; ID number CRD42022340968) of immune-related radiogenomic biomarkers in glioblastoma. The search strategy followed Preferred Reporting Items for Systematic Reviews and Meta-Analysis (PRISMA)[15] (Figure 1; Supplementary Table S1).

## Search Strategy and Selection Criteria

Search terms were applied to PubMed, MEDLINE, and EMBASE databases using medical subject headings (MeSH) terms[16] to identify original research articles published from January 1990 to January 2023 (Supplementary Table S2). A low-precision "high sensitivity search"[17] was conducted using subject headings and exploding terms. Studies not published in English,[18] editorials, conference proceedings, commentaries, letters, book chapters, laboratory-based or animal studies, preprints, or articles without peer review were excluded.

## Inclusion Criteria

The patients studied were adults aged over 18 diagnosed with glioblastoma. All studies with abstracts where MRI was used to develop and/or validate biomarkers of the tumor-host immune apparatus were included.

## Exclusion Criteria

All studies related to non-glial tumors; pediatric patients; vaccine trials; imaging other than MRI; and invasive studies including intratumoral injections or nanoparticle administration, were excluded.

## Appraisal of Quality

The Quality Assessment of Diagnostic Accuracy Studies 2 (QUADAS 2) tool[19] was used to assess the quality of the studies focusing on risk of bias and concerns regarding applicability. Relevant items from the Checklist for Artificial Intelligence in Medical Imaging (CLAIM) were also used to appraise studies[20] (Supplementary Table S3).

## Data Extraction

Data related to the type of study; MRI sequences; genomic markers; radiological markers, and their performance accuracy; and machine learning techniques employed, were extracted. Biomarkers were defined as diagnostic, prognostic, predictive, or monitoring according to the FDA-NIH BEST (Biomarkers, Endpoints, and other Tools) applied to neuro-oncology.[21]

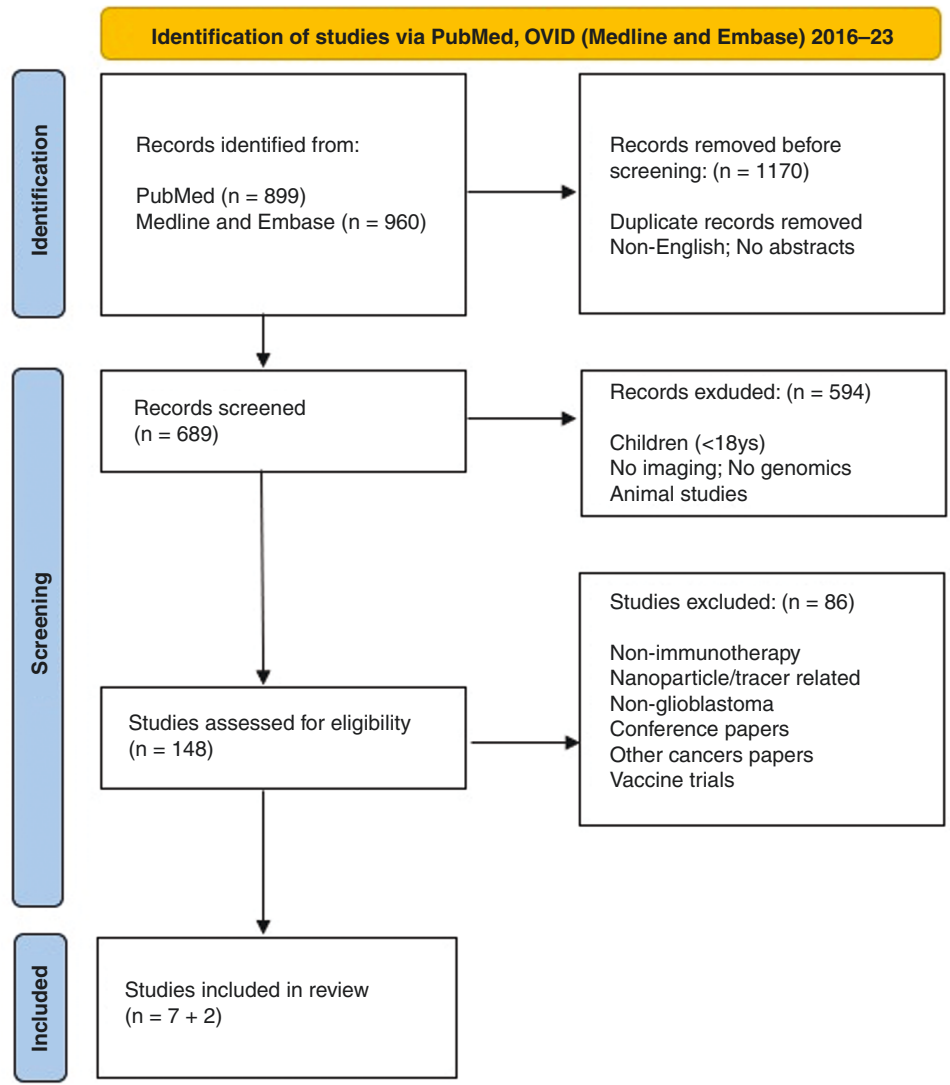

**Figure 1.**    Search strategy of systematic review for immune-related radiogenomic biomarkers in glioblastoma.

## Data Analysis

PG, a neurosurgeon with 6 years of clinical and research experience performed the literature search, which was independently reviewed by TB, a neuroradiologist with 15 years of clinical and research experience. Any discrepancies were resolved after discussion. A meta-analysis could not be performed due to a lack of sufficient homogenous data from the systematic review and marked heterogeneity in the methodology of studies.

## Results

Nine studies were included from 686 screened studies based on the PRISMA assessment (Figure 1). All studies[22–30] were retrospective and published after 2016 following the release of iRANO criteria for assessment of response to immunotherapy.[31] Seven studies were radiogenomic and were the focus of the systematic review to achieve the primary objective (Table 1). The remaining 2 were non-radiogenomic (Table 2) but included for illustrative purposes to highlight how researchers can develop immunotherapy biomarkers without any association with genomic information (secondary objective).

## Study Datasets

All studies included patients with histologically diagnosed "glioblastomas, isocitrate dehydrogenase (IDH)-wild type" or "astrocytoma, IDH-mutant, grade 4" who had undergone standard of care treatment.[6,34]

**Table 1.** Characteristics of *Radiogenomic* Studies Included in the Systematic Review

| Paper | Study design | Target condition | Dataset(s) | Available demographic information % male Age (mean ± SD) Ethnicity n (%) | Reference standard | Index test | Index test features selected | Type of test set | Test set performance |
|---|---|---|---|---|---|---|---|---|---|
| Cho HR et al., 2018[19] | Retrospective Single center | Myeloid cell marker expression level (CD11b, CD68, CSF1R, CD163, CD33, CD123, CD83, CD63, CD49d, CD117) Lymphoid cell marker expression level (CD4, Cd3e, CD25, CD8) | 60 patients Preoperative MRI:T2 FLAIR T1 CE, ADC, DSC Manual Segmentation of VOI | 58% male Mean age 54.2 ± 11.4 Data from Korea Ethnicity: N/A | RNA-level analysis using quantitative RT-PCR | CBV & ADC & tumor volume | nCBV within T2 FLAIR VOI (whole tumor) & T1 CE VOI (enhancing tumor) masks for CD68, CSF1R, CD33 & CD4. nCBV within T1 CE VOI (enhancing tumor) mask for CD11b. ADC within T2 FLAIR VOI (whole tumor) & T1 CE VOI (enhancing tumor) masks for CD49d & CD3e. ADC within T2 FLAIR VOI (whole tumor) mask for CD33 & CD123. ADC within T1 CE VOI (enhancing tumor) mask for CD25. FLAIR volume (whole tumor) & T1 CE VOI (enhancing tumor) mask for CD123, CD49d & CD117. | N/A (Train set only) | N/A (Train set only; index test features selected if *P* value < .05 with Pearson's correlation analysis) |

**Table 1.** Continued

| Paper | Study design | Target condition | Datasets(s) | Available demographic information % male Age (mean ± SD) Ethnicity n (%) | Reference standard | Index test | Index test features selected | Type of test set | Test set performance |
|---|---|---|---|---|---|---|---|---|---|
| Liao X et al., 2019[20] | Retrospective Multicenter | Prognosis: Survival data (< 1 year & > 1 year)—not immune-related Diagnosis: Immune-related & non-immune-related gene expression levels including: EREG, TIMP1, CHIT1, ROS1 | 137 patients—MRI data Preoperative MRI: T2 FLAIR MRI data: Training: 70% Test set: 30% Manual Segmentation of VOI 129 patients—Gene expression level and survival 46 patients -Intersection data (Gene expression level and MRI) | 63.2% male Mean age 61.7 ± 12.7 Data from US Ethnicity: N/A | Prognosis: Survival data (< 1 year and > 1 year) Diagnosis: RNA-level analysis using quantitative RT-PCR | 4 different models: GBDT, logistic regression, SVM, KNN | 72 radiomic features from T2 FLAIR MRI VOI selected for survival (first order and texture) | Prognosis: Internal hold out (72 radiomic features & survival; n = 129) Diagnosis: N/A (Train set only:72 radiomic features & gene expression level; n = 46) | Prognosis: GBDT accuracy = 0.81 AUC-ROC = 0.79 < 1 year; & = 0.81 > 1 year Diagnosis: N/A (Train set only: moderate (0.3 < |r| < 0.5) to high (|r| > 0.5) Pearson's correlation) |
| Jajamovich GH et al., 2016[22] | Retrospective Multicenter | Gene expression level signatures including dendritic cell maturation Molecular subtypes of glioblastoma | Gene data: 39% for 558 patients (TCGA) Matched preoperative MRI data from 560 patients (TCIA) Preoperative MRI: T1, T1 CE, DWI, ADC Semi-automatic segmentation of VOI (Growcut method) | Gene data: 39% male, average age: 58.3 ± 14.2 MRI data: 66% male Mean age: 60.7 ± 12.6 15 mesenchymal, 10 classical, 16 neural & 9 proneural glioblastoma patients with MRI images Data from US Ethnicity: N/A | RNA-level microarray analysis Molecular subtypes using nearest-centroid-based classification Gene signature subgroups based on k-means clustering | ADC (mean, SD, skewness, kurtosis, entropy) | Mean ADC values from T1 CE VOI (enhancing tumor) mask selected for (1) tumor molecular subtypes & (2) other subgroups of gene signature (including immune response related subgroup containing CD4, CD86, major histocompatibility complex class I and II). | N/A (Train set only) | N/A (Train set only: Neural vs non-neural subtypes (P = .02, AUC-ROC = 0.75, CI: 0.57–0.88) (Train set only: index test features selected if P value < .05 with Spearman's correlation analysis. Gave ρ = −0.51 for immune response-related subgroup.) |

**Table 1.** Continued

| Paper | Study design | Target condition | Dataset(s) | Available demographic information % male Age (mean ± SD) Ethnicity *n* (%) | Reference standard | Index test | Index test features selected | Type of test set | Test set performance |
|---|---|---|---|---|---|---|---|---|---|
| Liu D et al., 2023[23] | Retrospective Multicenter | Immune cell infiltration score | For the development and validation of immune cell infiltration score: (1) Discovery Set TCGA & GEO cohort (*n* = 400 patients; Gene expression level & survival) (2) Validation Set CGGA cohort (*n* = 374 patients; Gene expression level & survival) Matched dataset TCGA-TCIA and CPTAC cohorts (*n* = 70; Gene expression level & MRI and survival) Independent dataset NBH cohort (*n* = 149; MRI and survival) Preoperative MRI: T1 CE, T2 Manual segmentation of VOI | TCGA cohort: 64.8% male Mean age: 59.41 (SD 13.7) GEO cohort 58.5% male Mean age 52.06 (SD 13.6) CGGA cohort 60.4% male Mean age: 48.12 (SD 13.37) Matched cohort: 62.9% male Mean age: 60.34 (SD 12.67) NBH cohort: 59.1% male Mean age: 59.17 (SD 10.6) Data from US & China Ethnicity: N/A | Diagnosis: LM22 signature for immune cell expression using CIBERSORT. Immune & stromal content using ESTIMATE. Immune cell infiltration score = $\Sigma$PC1A −$\Sigma$PC1B Grouped as high or low immune cell infiltration score Prognosis: Survival data | SVM Model | 11 radiomic features from MRI T2 & T1 CE VOI were selected for immune cell infiltration grouping. (note: high interrater reliability of features was also a selection criterion) | Diagnosis: cross-validation only (11 radiomic features & immune cell infiltration group; *n* = 70) Prognosis: External hold-out (11 radiomic features & survival; *n* = 149) | Diagnosis: cross-validation only AUC-ROC = 0.96; accuracy = 0.94; recall = 0.91; F1 score = 0.93 Prognosis: Cox HR = 0.63 (0.46–0.88; *P* = .051) |

**Table 1.** Continued

| Paper | Study design | Target condition | Dataset(s) | Available demographic information % male Age (mean ± SD) Ethnicity n (%) | Reference standard | Index test | Index test features selected | Type of test set | Test set performance |
|---|---|---|---|---|---|---|---|---|---|
| Rao et al., 2016[24] | Retrospective Multicenter | Prognosis: Survival (median survival difference of 1 year)-not immune-related Diagnosis: gene expression levels (immune-related & non-immune-related; mRNA & miRNA expression data) | 92 patients: 44 patients-training set; 48 patients-test set Preoperative MRI; VASARI (Visually AcceSSAble Rembrandt Images) standardized feature set from T1, T2, T1 CE, T2 FLAIR MRI images No segmentation performed for the study | 65% male Age: N/A Data from US Ethnicity: N/A | Prognosis: Survival data (OS, PFS) Diagnosis: Normalized gene expression data from mRNA sequencing | k-adaptive partitioned VASARI features: volume-class, enhancement quality, proportion contrast-enhancing tumor, proportion noncontrast-enhancing tumor, proportion necrosis, proportion edema, T1/FLAIR envelope ratio, enhancing margin thickness, distribution, & hemorrhage | 3 radiomic features (volume-class, T1/FLAIR ratio, hemorrhage) selected for survival (>2: Group 1; ≤2: Group 0) | Prognosis: Internal hold-out (radiomic feature groups & survival PFS; n = 48) Diagnosis: N/A (Train set only: radiomic groups & gene expression level; n = 48) | Prognosis: High (median: 19 months) vs low survival (median: 7 months) group logrank test P = .0001 PFS P = .058 Diagnosis: N/A (Train set only: radiomic groups have differential gene expression; with 10 combinations representing immune-related & metabolic pathways when multiple testing corrections for significance with false discovery rate cut off 0.1) |
| Narang et al., 2017[25] | Retrospective Multicenter | Gene expression levels related to CD3 T cell activity | 79 patients (TCGA) used for biomarker development Preoperative MRI: T1 CE, T2 FLAIR Training and testing MD Anderson cohort: n = 69; 35 patients-training set; 34 patients-test set Semi-automatic segmentation of VOI | 52% male Mean age: 57.96 (SD 13.8) Data from US | Biomarker development: Normalized gene expression data for CD3E/D/G from mRNA sequencing Training and testing dataset: CD3 immunohistochemistry-derived cell counts | Symbolic regression model | 6 texture-based radiomic features (GLSZM, kurtosis, NGTDM) | Diagnosis: Internal hold-out (radiomic features & dichotomized CD3 counts (n = 34) | CD3 prediction model (using 6 features) AUC-ROC = 0.85 (95% CI: 0.66–0.94) Accuracy = 0.77 (95% CI: 0.59–0.89); recall = 0.73; F1 score = 0.80 |

**Table 1.** Continued

| Paper | Study design | Target condition | Dataset(s) | Available demographic information % male Age (mean ± SD) Ethnicity n (%) | Reference standard | Index test | Index test features selected | Type of test set | Test set performance |
|---|---|---|---|---|---|---|---|---|---|
| Hsu et al., 2020[28] | Retrospective Multicenter | Diagnosis: Gene Expression levels & corresponding immune signatures (CTL, aDC, Treg, MDSC) Prognosis: Survival (median OS) | 154 samples (TCGA RNAseq data) 116 patients (TCIA MRI data) Matched data for training (MRI & gene expression): 32 Independent Test data: 84 (MRI, survival; limited gene expression data) Preoperative MRI: T1 CE, DWI Manual segmentation of VOI | 60% male Median age: 62.5 (Q1–Q3: 52.3–70.3; training data); 59 (Q1–Q3: 52.0–66.0; test data) Data from US Ethnicity: N/A | Diagnosis: Gene set enrichment analysis to obtain enrichment levels of immune cells; categorized into immune signatures. Normalized gene expression data from mRNA sequencing & microarray Prognosis: Survival data (OS) | Logistic regression | 2–18 radiomics features: first order, GLRLM, GLCM from ADC (and T1 CE, from within T1 CE VOI data); selected for each immune signature. Also radiomics features: selected for survival group (median OS > 500 days) vs < 270 days survival). | Diagnosis: cross-validation only (radiomic features & immunophenotype groups; $n = 32$) Internal hold-out N/A (radiomic features & limited immunophenotype groups; $n = 84$) Prognosis: Internal hold-out N/A (T1 CE trained model radio-immune patient groups & survival data; $n = 84$) | Diagnosis: cross-validation only: CTL: Accuracy: 0.72 (T1 CE), 0.71 (ADC); AUC: 0.70 (T1 CE) aDC: Accuracy: 0.75 (T1 CE), 0.61 (ADC); AUC: 0.74 (T1 CE); 0.61 (ADC) Treg: Accuracy: 0.81 (T1 CE), 0.68 (ADC); AUC: 0.77 (T1 CE), 0.56 (ADC) MDSC: Accuracy: 0.88 (T1 CE), 0.79 (ADC); AUC: 0.85 (T1 CE), 0.70 (ADC) Prognosis: Internal hold out for comparison of immunophenotypes: for 2 selected types (median OS > 500 days vs < 270 days survival) group logrank test $P = .02$ |

TCGA, The Cancer Genome Atlas; ADC, apparent diffusion coefficient; IADC VOI, volume of interest with intermediate apparent diffusion coefficient; KNN, K nearest neighbors; PCA, principal component analysis; LDA, linear discriminant analysis; SVM, support vector machine; GBDT, gradient boost decision tree; MGMT, $O^6$-methylguanine–DNA methyltransferase; MRI, magnetic resonance imaging; TAMs, tumor-associated macrophages; AUC, area under the curve; T1, T1 weighted sequence; CE, contrast enhanced; T2, T2 weighted sequence; FLAIR, fluid attenuated inversion recovery; DWI, diffusion weighted imaging; RANO, response assessment in neuro-oncology criteria; ML, machine learning; DSC, dynamic susceptibility contrast; VOI, volume of interest; ROI, region of interest; nCBV, normalized relative CBV; OS, overall survival; PFS, progression-free survival; GLSZM, gray-level size zone matrix; NGTDM, neighborhood gray tone difference matrix; CTL, cytotoxic T lymphocytes; aDC, activated dendritic cells; MDSC, myeloid-derived suppressor cells; Treg, T regulatory cells.

**Table 2.** *Non-Radiogenomic* Studies in the Review

| Paper | Study Design | Target condition | Dataset(s) | Available demographic information % male Age (mean ± SD) Ethnicity *n* (%) | Reference standard | Index test | Index test features selected | Type of test set | Test set performance |
|---|---|---|---|---|---|---|---|---|---|
| George E et al., 2022[18] | Retrospective Multicenter | OS PFS | 113 patients from PD-L1 inhibition immunotherapy trial (NCT02336165) pretreatment & 8-week posttreatment MRIs: T1, T1 CE, T2, T2 FLAIR Manual segmentation of VOI 3 train-test combinations giving range (n-n): Train = 60–74 Test = 29–43 | 69% 55.2 ± 11.5 years Data from US Ethnicity: White 99 (87.6%) African American 1 (0.9%) Asian 1 (0.9%) Other 3 (2.7%) Unknown 9 (8.0%) | Survival data PFS derived from modified RANO[32] (unclear number of readers & seniority) | Random Forest model | Radiomics shape & texture extracted from T2 FLAIR VOI (whole tumor) & T1 CE VOI (enhancing tumor) masks. Note: only features within top 20 in all 3 test sets for OS: GLCM correlation T1 & maximal axial diameter in T1 CE VOI. Note: only feature within top 20 in all 3 test sets for PFS: GLRLM RLVT2 in T2 FLAIR VOI. | External (3 combinations of sites—similar to cross-validation methodology) | AUC pretreatment OS 0.47–0.52 PFS 0.47–0.52 AUC Posttreatment OS 0.69–0.75 PFS 0.68–0.71 |
| Qin L et al., 2017[21] | Retrospective Multicenter | Anti-PD-1 +/− Anti-CTLA-4 therapy response in recurrent glioblastoma | 10 immunotherapy trial patients (NCT02017717; NCT02054806) (5 benefit group and 5 no-benefit group) Postoperative MRI: T1, T1C, T2 FLAIR, ADC Manual segmentation of VOI | Trial data from respective trials | Survival data (< 5 months & > 5 months) Note: without unequivocal imaging, clinical, or histopathologic evidence of progression RANO[33] measures on T1 CE image | Intermediate ADC volume (IADC VOI) | IADC VOI change from FLAIR VOI | N/A (Train set only) | N/A (Train set only; 100% accuracy) |

Among the radiogenomic studies, 6/7 (85.7%) were multicenter and one was performed using a dataset of 60 consecutive patients from a single center.[23] The Cancer Imaging Archive (TCIA) MRI data (https://www.cancerimagingarchive.net/collection/tcga-gbm/) and corresponding genomic data from The Cancer Genome Atlas were used as datasets in all the multicenter studies.[24,26–30] In one TCIA-TCGA study, Liao et al.[24] developed radiomic biomarkers corresponding to immune-related gene expression.[35–37] The study included 137 patients with TCIA MRIs, of which 46 had corresponding genomic information. In a second study, Jajamovich et al.[26] developed imaging biomarkers from 558 patients with TCGA genomic information, of which 50 had corresponding MRIs. In a third study, Liu et al.[27] used multiple datasets (TCGA, Chinese Glioma Genome Atlas, and Clinical Proteomic Tumor Analysis Consortium RNA-sequencing data; GSE13041 and GSE83300 RNA microarray data; TCIA and local institution imaging data) and developed biomarkers using a cohort of 774 patients with mRNA gene expression data from multicenter datasets including 70 patients matched with MRI and mRNA data (TCGA, Clinical Proteomic Tumor Analysis Consortium). Subsequently, the biomarkers were validated using MRI and survival data from a third independent cohort of 149 patients from a single center. In the fourth study, Rao et al.[28] studied 92 patients from the TCGA database with MRI, mRNA, miRNA, and survival data. In the fifth study, Narang et al.[29] developed biomarkers using matched MRI and mRNA data from 79 patients within the TCIA-TCGA database. The biomarkers were then trained on 35 patients and tested on 34 patients from a separate hospital cohort. Hsu et al.[30] identified biomarkers using matched MRI and mRNA data of 32 patients from TCIA-TCGA database and tested them on 84 patients with MRI and survival data from the TCIA database; limited mRNA data were also available in the test set.

Out of the 2 non-radiographic studies, one analyzed recurrent tumors[25] and the other included a mixture of newly diagnosed and recurrent tumors.[22] Both studies included patients from immunotherapy clinical trials.[22,25] Specifically, George et al.[22] used data from a multicenter phase II programmed death-ligand 1 clinical trial (NCT02336165) with a sample size of 113 patients partitioned into training and test sets. In the second study, Qin et al.[25] studied 10 consecutive patients enrolled in clinical trials of anti-PD-1 therapy with or without anti-CTLA-4 therapy (NCT02017717; NCT02054806).

## Magnetic Resonance Imaging

The images used to develop biomarkers were obtained from either $T_1$-weighted (T1), $T_1$-weighted contrast-enhanced (T1 CE), $T_2$-weighted (T2), $T_2$-weighted Fluid Attenuated Inversion Recovery (T2 FLAIR), dynamic susceptibility contrast-enhanced (DSC) sequences or diffusion-weighted imaging/apparent diffusion coefficient maps (DWI/ADC). All radiogenomic studies included either T2 FLAIR (4/7, 57.1%) or T1 CE (6/7, 85.7%) images as a minimum.

## Machine Learning, Radiomics, and Statistical Analysis

Eight studies (8/9; 88.9%) used manual or semi-automated segmentation for determining the image volume of interest and classified extracted image features with classical machine learning or advanced statistical modeling techniques while one study[28] did not use segmentation and applied VASARI (Visually AcceSAble Rembrandt Images) standardized features to advanced statistical modeling techniques. No deep-learning techniques were used. The extracted image features were either radiomic-based and obtained from structural images or consisted of quantitative ADC metrics. An exception was one study, which also extracted cerebral blood volume metrics in addition to ADC metrics.[23] Radiomic features were extracted using Pyradiomics[24,27] (https://github.com/AIM-Harvard/pyradiomics) or the open source radiomics package by Vallières[22] (https://github.com/mvallieres/radiomics).

All radiogenomic studies[23,24,26–30] (7/7, 100%) developed diagnostic imaging biomarkers that identified glioblastoma with immune-related gene signatures,[24,26–30] immune cell markers[23] or immune infiltration scores.[27] Four radiogenomic studies (4/7; 57.1%)[24,27,28,30] also demonstrated that the imaging biomarkers were prognostic by correlating imaging features with survival. The 2 non-radiographic studies developed a prognostic[22] biomarker related to survival, and a predictive imaging biomarker[25] that correlated with immunotherapy-related treatment response, respectively.

The radiogenomic studies[23,24,26–30] (7/7, 100%) developed biomarkers by correlating MRI features with immune-related gene expression levels[23,26,28–30] (diagnostic biomarkers), composite scores derived from them called "immune cell infiltration scores"[27] (diagnostic biomarkers) or survival data[24,27,28,30] (prognostic biomarkers). In 5/7 (71.4%) studies[23,24,27,28,30] indirect methods were used to determine that the imaging biomarkers were clinically meaningful (Table 3) by correlating the classified groups of (1) an imaging-based survival classifier with immune-related gene expression levels,[24,28,30] or (2) an imaging-based immune-related gene expression level classifier with progression-free survival,[23] or (3) an imaging-based immune cell infiltration classifier with survival.[27]

Cho et al.[23] compared MRI-derived ADC and normalized relative cerebral blood volume (nCBV) values with lymphoid and myeloid cell marker expression levels, demonstrating that CD68 (tumor-associated macrophages; TAMs), CSF1R (TAMs), CD33 (myeloid-derived suppressor cell) and CD4 (regulatory T-cell) levels positively correlate with nCBV values; and CD3e (cytotoxic T-cell) and CD49d (bone marrow-derived cells) negatively correlate with ADC values. These findings persisted regardless of whether enhancing tumor or whole tumor was analyzed. CD123 (dendritic cells), CD49d, and CD117 (mast cells) were also negatively correlated with tumor volume. To determine if the immune cell markers selected in the study were clinically meaningful, a Cox proportional hazard analysis of progression-free survival was performed with only CD49d expression proving significant.

Liao et al.[24] used Pyradiomics to extract shape, first order, and texture-based radiomic features from 2D FLAIR images, and employed 4 different models on the data, namely Gradient Boosting Decision Tree (GBDT), logistic regression, support vector machine (SVM) and k-nearest neighbors (KNN). They showed that GBDT performance was best among the 4 models with an accuracy of 0.81 for classifying images into those related to short or long survivors. Six gene expression levels differed between the 2 survivor classes, 3 of which were moderately highly correlated with the most discriminative radiomic features. These 3 genes were tissue inhibitors of metalloproteinases 1 (TIMP1), repressor of silencing 1, and epiregulin (EREG), all of which have immune-related functions.[35–37]

Using a different approach, Jajamovich et al.[26] used MRI-derived ADC correlation analysis on gene expression data grouped into molecular subtypes as well as gene subgroups. The researchers demonstrated a negative correlation of mean ADC values with an immune-related gene signature subgroup containing CD4, CD86, and major histocompatibility complex class I and II which are associated with dendritic cell maturation, the complement system, and macrophage function.

Liu et al.,[27] refined gene expression grouping further still using extracted shape, first order, wavelet, and texture-based radiomic features from intra- and peri-tumoral regions. Key features were selected using recursive feature elimination and SVM to generate a predictive model that classified tumors into those with low or high immune cell infiltration scores. These immune cell infiltration scores represented those immune cell infiltration patterns in the gene expression data that persisted in different datasets and were shown to be prognostic for survival. In an independent MRI dataset, the SVM model classified patients into predicted classes of low and high immune cell infiltration; only survival data was available as a reference standard.

**Table 3.**   Radiological Biomarkers Identified in the Systematic Review

| Paper | Radiogenomic biomarker | Radiological biomarker | Prognostic | Monitoring | Diagnostic (Radiogenomic component) | Predictive | Direct method of biomarker development | Indirect methods of biomarker development |
|---|---|---|---|---|---|---|---|---|
| Cho et al [19] | Yes | Quantitative MRI features (ADC, nCBV, tumor volume) | N/A | N/A | Yes (Expression levels of immune cell markers) | N/A | Correlation of ADC, nCBV, tumor volume features & expression levels of immune cell markers | Expression levels of immune cell markers & PFS (without images) Note: PFS derived from RANO[32] |
| Liao et al [20] | Yes (secondary) | Radiomic MRI features (T2 FLAIR) | Yes (Radiomics and survival—not immune-related) | N/A | Yes (Indirect: expression levels of immune-related genes and radiomics) | N/A | Using survival data of patients to classify radiomic features -not immune-related | Expression levels of the genes identified (all immune-related) distinguished 2 survival groups. These immune-related gene expression levels were then compared to radiomics. |
| Jajamovich et al [22] | Yes | Quantitative MRI features (ADC) | N/A | N/A | Yes (tumor subtype. Immune-related gene expression signature subgroup) | N/A | Correlation of ADC features and (1) tumor subtype and (2) immune-related gene expression subgroup | N/A |
| Liu et al [23] | Yes | Radiomics MRI features (T2 and T1 CE) | Yes (Radiomics and survival) | N/A | Yes (Immune cell infiltration score and radiomics) | N/A | Using radiomic features to classify patients into groups with high or low immune cell infiltration scores | Assessment of prognosis by (1) correlating immune cell infiltration scores with survival in 2 datasets, then, (2) correlating radiomics with immune cell infiltration scores in a third dataset (correlation of immune cell infiltration scores and survival endured), then, and (3) correlating radiomics and survival in fourth dataset |
| Rao et al [24] | Yes (secondary) | Quantitative and qualitative MRI features (VASARI) | Yes (MRI features and survival-not immune-related) | N/A | Yes (Expression level of immune-related genes and MRI features) | N/A | Using survival data to classify radiomic feature groups-not immune-related | Expression level of genes identified (immune-related and non-immune-related) that distinguished 2 radiomic feature group |
| Narang et al [25] | Yes | Radiomics MRI features (T1 CE, T2 FLAIR) | N/A | N/A | Yes (Gene expression levels related to CD3T cells) | N/A | Correlation of radiomics features with gene expression signature for CD3T cells | N/A |
| Hsu et al [26] | Yes | Radiomics MRI features (T1 CE, ADC) | Yes (radio-immune patient groups and survival) | N/A | Yes (immune gene signatures and radiomics) | N/A | Using radiomic features to classify patients into groups with different gene expression immune signatures | Predicted immunophenotype patient groups with significant differences in median overall survival |
| George et al [18] | No | Radiomics features (VOI structural images) | Yes (Radiomics and survival) | N/A | N/A | N/A | Using survival data of patients in immunotherapy clinical trials to regress radiomic features | N/A |
| Qin et al [21] | No | Quantitative MRI features (IADC VOI) | N/A | Yes (longitudinal change in IADC VOI during immunotherapy) | N/A | Yes (IADC VOI change and therapeutic benefit) | Using survival data of patients in immunotherapy clinical trials to determine feature change longitudinally | (RANO[33] but minimally supportive) |

VOI, volume of interest; N/A, not applicable; ADC, apparent diffusion coefficient; IADC VOI, volume of interest with intermediate apparent diffusion coefficient; IADC VOI, volume of interest with intermediate apparent diffusion coefficient; OS, overall survival; PFS, progression-free survival; VASARI, Visually AcceSAble Rembrandt Image; SVM, support vector machine; RF: Random Forest; GBDT: Gradient boom decision tree.

Rao et al.[28] used MRI VASARI features to dichotomize the data into 2 groups with corresponding scores according to the tumor volume class, T1/FLAIR ratio, and hemorrhage values. These radiomic groups were prognostic for survival and showed significant differences in gene expression levels within immune-related pathways (inducible co-stimulator (iCOS-iCOSL) signaling in T helper cells; retinoid X receptor (RXR) activation; and phosphoinositide 3-kinase (PI3K) signaling in B lymphocytes).

Narang et al.[29] obtained 6 radiomic-based imaging features (Gray-Level Size Zone Matrix, kurtosis, Neighborhood Gray Tone Difference Matrix) after feature selection tailored to gene expression levels of CD3T cells using the Boruta algorithm. Using dichotomized CD3 counts, they trained and tested the classifier using the 6 features. A multivariate regression analysis demonstrated that the classifier was not confounded by clinical factors or tumor volume.

Hsu et al.[30] identified radiomic-based imaging features related to T1 CE and ADC images (first order, gray-level run-length matrix, gray-level co-occurrence matrix (GLCM)) that were able to classify clustering-derived immune cell subset patient groups based on immune profile combinations (cytotoxic T lymphocytes (CTLs), activated dendritic cells (aDCs), T regulatory cells (Tregs), myeloid-derived suppressor cells) using logistic regression models. The features were selected using random forest and information gain algorithms.

### Radiological Imaging Biomarker Summary

Biomarkers extracted from MRI volumes of interest that correlated with various immune-related markers in patients with glioblastoma included ADC values, nCBV values, and image-based (VASARI, radiomics) features.

ADC biomarkers were negatively correlated with, firstly, CD3e and CD49d expression levels and, secondly, an immune-related gene signature (CD4, CD86, major histocompatibility complex class I and II) in respective studies.[23,26] Similarly, nCBV biomarkers were positively correlated with expression levels of CD68, CSF1R, CD33 and CD4.[23] Radiomic biomarkers (shape, first order, wavelet, and texture) were predictive of firstly, immune infiltration patterns/scores or CD3 expression levels in respective studies,[27,29] or secondly survival, which was shown to be correlated with immune-related genes (TIMP1, repressor of silencing 1, EREG), immune cell infiltration scores or other immune signatures, in respective studies.[24,27,30] Simpler radiomic biomarkers (tumor volume-class, T1/FLAIR ratio, and hemorrhage phenotype) were predictive of survival, which was shown to be correlated with immune-related pathways (iCOS-iCOSL, RXR, and PI3K).[28] Similarly, tumor volume was negatively correlated with CD123, CD49d, and CD117.[23] The immune-related genomic and corresponding radiological biomarkers identified in this review are summarized in Table 4.

### Bias Assessment and Applicability Concerns

A qualitative analysis of the risk of bias and concerns regarding applicability was performed for each study and is summarized in Supplementary Figure S1. Six (6/9; 67%) studies had a high risk of index test bias. The risk of bias was high or unclear in 6/9 (67%) studies regarding patient selection and was unclear in 4/9 (44%) studies regarding the reference standard used. Concerns of study applicability were high regarding the index test in 6/9 (67%) studies, high or unclear regarding patient selection in 7/9 (78%) studies, and unclear regarding the reference standard used in 5/9 (56%) studies.

## Discussion

### Summary of Findings

The systematic review demonstrated that radiological biomarkers, namely ADC values, nCBV values, and radiomic features (VASARI, texture, shape, histogram, and wavelet) extracted from different MRI sequences, correlated with immune-related genetic markers and were developed as noninvasive radiogenomic biomarkers. Some studies used internal hold-out datasets for analytical biomarker validation[21]; however, none used external hold-out datasets to validate the trained biomarker. Some non-radiogenomic biomarkers (ie, without any correlation with immune-related genetic markers) were developed to predict response to immunotherapy. All reviewed studies are best considered as "proof of concept."

### Limitations

#### Studies Assessed

All studies employed retrospective designs. Limitations encompassed 6 main areas.

First, differences in the type of genomic data (single vs bulk RNA-sequencing data; microarray data, polymerase chain reaction or immunohistochemistry-based data) and their harmonization in each study confound pooled inferences from the different studies (Supplementary Table S4).

Second, patients underwent MRI imaging in different centers where there were differences in scanner manufacturer and local MRI sequence protocols. Different postprocessing steps were deployed in each study to tackle these differences but lack uniformity (Supplemental Table S5). It is plausible that there could be subsequent variability in the imaging features between centers confounding pooled inferences from the different studies.

Third, patient selection for the majority of studies was based on what had been included in public datasets (especially TCIA/TCGA) or small sets of local hospital data. Not only did the sample appear to be similar or the same in almost all studies (from TCIA/TCGA), but there was no clear and detailed explanation regarding the process of patient selection. For example, there was no clarity regarding patient selection being continuous or at random. Furthermore, other eligibility criteria varied amongst all the studies and again the details were unclear in the majority of studies. Confounded patient selection may mean that the study samples are not representative of the intended population ("glioblastomas, isocitrate dehydrogenase (IDH)-wild type" and "astrocytoma, IDH-mutant, grade 4") which limits the generalizability of the results

**Table 4.** Immune-Related Genomic Biomarkers With Corresponding Radiological Biomarkers Identified in the Review

| Paper | Immune-related genomic biomarker/s | Immune Function/ associated immune cells | Clinical status | Radiological biomarker | Radiological status |
|---|---|---|---|---|---|
| Cho et al.[23] | CD68, CSF1R, CD33, CD4, CD49d, CD11b, CD123, CD25, CD117 | CD68,[38] CSF1R[38,39]:TAMs CD33[40]: myeloid-derived suppressor cells (MDSCs) CD4[41–43]: helper T cells, cytotoxic T cells CD3e[44,45]: helper T cells, cytotoxic T cells CD49d[46]: myeloid cells CD11b[47]: macrophages, neutrophils, NK cells, memory B cells, cytotoxic T cells CD123[48]:dendritic cells CD25[49]:T cells, B cells, NK cells, regulatory T cells CD117[50]: hematopoietic stem and progenitor cells, pro-B cells, pro-T cells | Established | Quantitative MRI features (ADC, nCBV, T2 FLAIR) | Not established |
| Liao et al.[24] | TIMP1, ROS1, EREG, and CHIT1 | TIMP1[51]: Dendritic cells, macrophage, neutrophils ROS1[52]: Under research EREG[53]: Influence expression of PD-L1 CHIT1[54]: macrophages | Not established | Radiomic MRI features (T2 FLAIR) | Not established |
| Jajamovich et al.[26] | CD4, CD86, MHC I, and MHC II | CD4,[41–43] CD86,[55] MHC I,[56,57] MHC II[58,59]: Dendritic cell maturation, TREM1 signaling, communication between innate and adaptive immune cells, production of nitric oxide and reactive oxygen species in macrophages, complement system | Established | Quantitative MRI features (ADC) | Not established |
| Liu et al.[27] | Immune cell infiltration score (low vs high) | N/A | N/A | Radiomic MRI features (T2, T1 CE) | Not established |
| Rao et al.[28] | Immune-related pathways derived from gene expression levels | N/A | N/A | Quantitative and qualitative MRI features (VASARI; T1, T2, T1 CE, T2 FLAIR) | Not established |
| Narang et al.[29] | CD3 | CD3[60]: Helper T cells, cytotoxic T cells | Established | Radiomic MRI features (T1 CE, T2 FLAIR) | Not established |
| Hsu et al.[30] | Enrichment-based Immune phenotypes based on cytotoxic T cells, activated dendritic cells, regulatory T cells (Tregs), myeloid-derived suppressor cells (MDSCs) | N/A | N/A | Radiomic MRI features (T1 CE, ADC) | Not established |
| George et al.[22] | Response to anti-PD-L1 immunotherapy | PD-L1[61,62]:T cells | Established | Radiomic MRI features (T1, T1 CE, T2 FLAIR) | Not established |
| Qin et al.[25] | Response to Anti-PD-1 +/− CTLA-4 immunotherapy | PD-1,[61,62] CTLA-4[63]:T cells | Established | Quantitative MRI features (ADC, T2 FLAIR) | Not established |

VOI, volume of interest; CD, cluster of differentiation; TAMs, tumor-associated macrophages; MHC, major histocompatibility complex.
Regarding the clinical status of the immune markers, we define "established" and "not established" arbitrarily as being established as an immune cell surface markers and vice-versa; regarding radiological status, "established" and "not established" as the radiological markers are clinically used as a biomarker for immune status and not radiologically established marker for immune status respectively.

to the clinic. It is noteworthy that even if generalizable to the pooled grade 4 gliomas, the biomarkers developed in the studies have not been optimized for IDH-wild-type glioblastoma alone (as the datasets preceded the 2021 WHO classification).

Fourth, details regarding the reference standards used in the majority of studies were unclear and it would be challenging to reproduce them. Furthermore, tumor heterogeneity within the TME is likely to confound reference standards and may be a limitation in all the studies as the biopsy sample of the tumor, and subsequent tumor-tissue genomic data, may not entirely represent the overall TME of the tumor.[64,65] The majority of the studies[23,24,26–28] have not addressed other confounding variables such as age at diagnosis, resection status (biopsy, subtotal resection, total resection), postsurgical treatment (complete/

incomplete Stupp protocol) and second-line treatment including immunotherapy that are likely to influence the development and validation of prognostic biomarkers.[24,27,28] Moreover, diagnostic biomarkers can also be confounded by the unique interaction between the central nervous system, immune system, and advanced age in patients with glioma.[66] An example relevant to 2 of the included studies[23,26] is that microglia express higher basal levels of MHCII and CD11b with age.[67]

Fifth, the variable index tests developed as radiogenomic biomarkers did not undergo rigorous analytical validation and none were clinically validated.[21] Internal hold-out test sets were used effectively to validate prognostic biomarkers after training in 2 studies[24,28] and a diagnostic biomarker in one study[29] (none were temporal hold-out test sets). Overall, these findings limit the generalizability of the results to the clinic.

Sixth, most studies employed indirect methods for biomarker development and validation. For example, an imaging biomarker might predict a gene expression signature; a separate dataset containing no imaging data might show that the same gene expression signature can predict survival. The separate dataset is not a hold test set for validating an imaging biomarker for either a gene expression signature or survival. The limitation is that such indirect methodology for imaging biomarker development shows there is some clinical relevance, but this is not analytical validation.[21] Most studies likely employed such methods as there are few datasets containing imaging data that is matched with gene expression (for diagnostic biomarkers) or survival (for prognostic biomarkers).

*Review Process*
Pooled diffuse glioma (WHO grades 2–4) studies were excluded from the review process as it was beyond the research question, but we acknowledge that the biomarkers obtained in these studies might be of use in glioblastoma.[68,69]

Publication bias may have affected the range of performance accuracy of the biomarkers included in this systematic review. The potential for publication bias may be heightened by the omission of preprints and materials that have not undergone peer review. This is particularly relevant in the data science community, where the rapid pace of development often outstrips the slower process of peer review, leading some researchers to avoid submitting their work to peer-reviewed journals.[17] The composition of the research team could therefore influence this bias. Teams with a stronger clinical focus might be more likely to seek publication in peer-reviewed journals, whereas those with a stronger emphasis on data science might not.

**Study Explanations and Relevance From a National and International Perspective**

The focus of most of these studies was on prognosis which may be of limited relevance to either identifying immune-related targets for immunotherapy; or for predicting therapeutic response to immunotherapy. Novel immunotherapeutic approaches are currently being explored for glioblastoma but the translational landscape from basic scientific evidence to efficacious clinical treatment is still far behind other cancers.[9,70–80] Two areas of research can be combined to help develop panels of biomarkers which may be useful to stratify immunotherapy to treat particular tumors, and thereby contribute meaningfully to translation. First, studies focusing on immune-related genes and the immune tumor microenvironment (TME) in glioblastoma as well as melanoma, ovarian, lung, and colon cancers have demonstrated potential immunotherapy targets and therefore desirable prediction classes for radiogenomic analysis.[9,35–37,64–66,72–76,81–85] Second, there is an expanding arsenal of techniques to extract features including radiomics and deep learning features that can be used to develop imaging biomarkers in glioblastoma,[86–91] and even a decade ago non-immune radiogenomic glioblastoma studies demonstrated considerable promise.[92] It is plausible that these 2 advancements, alongside an expanding number of new data repositories, may lead to the development of important biomarkers and allow translation to succeed—the review shows we are currently at a proof-of-concept stage.

**Current Evidence in the Field**

This is the first systematic review of immune-related radiogenomic biomarker studies for glioblastoma. One study that did not focus on glioblastoma patients but also included oligodendroglioma and astrocytoma patients, developed an immune TME radiomic signature.[93] Here it was shown that the heterogeneity of the immune TME harbors prognostic impact. Other studies of interest have used different modalities. Nagle et al.[94] demonstrated imaging biomarkers for labeled CD8 T cells using positron emission tomography (PET) imaging in glioblastoma mouse models and showed the ability to quantify CD8 T cells noninvasively. Similarly, various radiomic signatures associated with CD8 T cells were identified in a systematic review by Ramlee et al.[95] related to various cancers including glioma (high and low-grade), gastrointestinal cancer, head and neck cancer, hepatobiliary cancer, lung cancer, breast cancer, and melanoma and their respective CD8 T-cell-related radiomic signature obtained from imaging modalities such as PET, CT, and MRI.

Large high-quality multicenter studies are possible and should be the standard to aim for in neuro-oncology. In other oncology disciplines, this has been demonstrated. For example, Sun et al.[96] developed and validated CT-derived radiomic biomarkers related to tumor-infiltrating CD8 T cells in patients included in phase I trials of anti-programmed cell death protein-1 (PD-1) or anti-programmed cell death ligand 1 (PD-L1) monotherapy for solid malignant tumors. Similarly, Trebeschi et al.[97] developed CT-derived radiomic biomarkers for predicting response to immunotherapy in advanced melanoma and lung cancer patients. It is also noteworthy that platforms such as ImaGene (https://github.com/skr1/Imagene) have demonstrated the potential for reproducibility of radiogenomic analysis with initial feasibility experiments analyzing invasive breast carcinoma, and head and neck squamous cell carcinoma.[98]

## Implications for Future Research and Clinical Practice

The present review has revealed an absence of high-quality studies regarding immune-related radiogenomic markers in glioblastoma with concerns regarding bias and generalizability. Future large, multicenter, prospective studies using radiomic or deep learning methods are required for the development and validation of pertinent biomarkers. It is plausible that features extracted from images of modalities such as advanced MRI (including permeability, perfusion, diffusion, chemical exchange saturation transfer), MR spectroscopic imaging, and PET might provide additional information on tumor biology and microenvironment. Future studies could also develop and validate biomarkers for either IDH-wild-type glioblastoma alone which likely has a unique immune TME (biomarkers for postbiopsy settings at recurrence or during immunotherapy treatment),[99] or for lesions that are suspected to be glioblastoma (biomarkers for prebiopsy and neo-adjuvant settings which might include enhancing lower grade gliomas and other mimics). Candidate biomarkers need to be clinically validated in the setting of prospective studies. Whether a clinically validated biomarker demonstrates impact when used in conjunction with an intervention would require the biomarker to be integrated into immunotherapy clinical trials such as the CheckMate 143 study.[10] Even if prospective biomarker studies are clinically validated soon, for example, to provide a panel of diagnostic biomarkers ready for patient stratification in downstream research, the scarce level 1 evidence for immunotherapy benefit currently means that biomarker studies demonstrating impact (ie, validated predictive biomarkers) when used in conjunction with an intervention, are unlikely to emerge soon.

Future studies might also use spatial transcriptomics or single-cell sequencing to better understand the role of immune cells in disease progression and lead to the discovery of new classes for radiogenomic analysis. Ultimately, there is the potential to produce noninvasive imaging biomarkers for neo-adjuvant immunotherapy stratification as part of personalized medicine within the next decade.

## Supplementary material

Supplementary material is available online at *Neuro-Oncology* (https://academic.oup.com/neuro-oncology).

## Keywords

deep learning | glioblastoma | immunotherapy | machine learning | radiogenomics

## Funding

TCB supported by the Wellcome Trust [WT 203148/Z/16/Z]. PG is supported by King's College London postgraduate research (PGR) international studentship.

## Acknowledgment

We thank all staff supporting the programme of work from (1) the KCL School of Biomedical Engineering & Imaging Sciences in particular, Giusi Manfredi, Vicky Goh, Patrick Wong, Denise Barton, Valentina Vitiello and Sebastien Ourselin, as well as (2) from King's College Hospital NHS Foundation Trust, in particular, Ann-Marie Murtagh, Jasmine Palmer and all others in R & D. We thank Dr James Arnold for reviewing the manuscript.

## Conflict of interest statement

None.

## Authorship statement

Study concept and design: P.G., M.M., and T.C.B.; Literature review and qualitative analysis: P.G. and T.C.B.; Writing the manuscript: P.G., M.M., and T.C.B.. All authors reviewed and approved the final manuscript.

## Data availability

All data has been made available in the supplemental file.

## Affiliations

School of Biomedical Engineering & Imaging Sciences, King's College London, London, UK (P.G., M.M., T.C.B.); Department of Neurosurgery, Kings College Hospital NHS Foundation Trust, London, UK (P.G., K.A.); Department of Oncology, University College London, London, UK (B.K.); Genomics England, London, UK (G.K., P.A.); Division of Genetics and Epidemiology, The Institute of Cancer Research, Sutton, UK (R.H.)